\documentclass[prl,aps,twocolumn]{revtex4}
\usepackage{epsfig}
\begin{document}
\title{Structural transformation induced by magnetic field and colossal
magnetoresistance response above 313 K in MnAs}
\author{J. Mira,$^1$ F. Rivadulla,$^{2,}$\footnote{On leave from 
Departamento de
Qu\'\i{}mica-F\'\i{}sica, Universidade de Santiago de Compostela, E-15782
Santiago de Compostela, Spain} J. Rivas,$^1$ A. Fondado,$^1$ R. Caciuffo,$^3$
F. Carsughi,$^3$ T. Guidi,$^3$ and J. B. Goodenough$^2$}
\address{$^1$Departamento de F\'\i{}sica Aplicada, Universidade de
Santiago de Compostela, E-15782 Santiago de Compostela, Spain
\\
$^2$Texas Materials Institute, ETC 9.102,
The University of Texas at Austin, Austin, Texas 78712, USA \\
$^3$Istituto Nazionale per la Fisica della Materia, Dipartimento di Fisica
ed Ingegneria dei Materiali e del Territorio, Universit\'a di Ancona,
I-60131 Ancona, Italy}

\begin{abstract}
MnAs exhibits a first-order phase transition from a ferromagnetic, 
high-spin metal
NiAs-type hexagonal phase to a paramagnetic, lower-spin insulator
MnP-type orthorhombic phase at $T_{C} = 313$ K. Here, we report the results
of neutron diffraction experiments showing that an external magnetic
field, $B$,
stabilizes the hexagonal metallic phase above $T_{C}$.
The phase transformation is reversible and constitutes the first 
demonstration of a
bond-breaking transition induced by a magnetic field. At 322 K
the hexagonal structure is restored for $B~>~4$ tesla.
The field-induced phase transition is accompanied
by an enhanced magnetoresistance of about $17~\%$ at $310$ K. We
discuss the origin of this phenomenon, which appears to be similar to
that of the colossal magnetoresistance
response observed in some members of the
manganese perovskite family.
\end{abstract}


\maketitle

MnAs is a commercially available material,
intensively studied, both theoretically and experimentally, since the
beginning of the last
century. Interest in this compound could come up again as a consequence of
new ideas and conjectures formulated during the last decade in
connection with the study of the
colossal magnetoresistance (CMR) response in Mn perovskites \cite{Coey}.
Among these
ideas is the invocation of a phase separation scenario for CMR manganese
oxides and related materials \cite{Moreo1}, that might be of 
particular relevance
in systems, like MnAs, where first-order phase transitions occur.

\begin{figure}
\epsfig{file=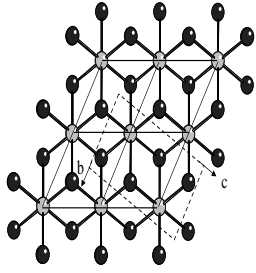,height=6cm,width=6cm}
\caption
{The orthorhombic unit cell of the $B31$ form of MnAs in relation to the
hexagonal $B8_{1}$ type cell. Mn (white circles) and As (black circles) atoms
are in the 4c special positions of the Pnma space group, (x,1/4,z), with
x = -0.005 and z = 0.223 for Mn, x = 0.275 and z = -0.082 for As.}
\label{un}
\end{figure}

To our knowledge,
MnAs was first studied by Heusler \cite{Heusler}, and later by Hilpert and
Dieckmann \cite{Hilpert1,Hilpert2,Hilpert3}, who discovered that the 
compound is
ferromagnetic with a Curie temperature of $T_C \simeq$ 313 K. Serres
\cite{Serres} and Guillaud \cite{Guillaud} found that a metal-insulator
transition occurs at $T_C$. The abrupt loss of magnetization at $T_C$ led Meyer
and Taglang \cite{Meyer} to assume a ferromagnetic-antiferromagnetic transition
at $T_C$. However, a neutron-diffraction study by Bacon and Street \cite{Bacon}
showed that this was not the case. At the same time, Willis and Rooksby
\cite{Willis} measured a large (1.86$\%$) discontinuous density change at $T_C$
that had been previously detected by Smits, Gerding, and Ver Mast \cite{Smits}.
Bean and Rodbell \cite{Bean} used the volume change to describe the first-order
loss of magnetization, which takes place with a latent heat of 1.79 cal/K
\cite{Bates1}, in terms of a volume-dependent exchange interaction.
De Blois and Rodbell \cite{deBlois} explored the
change in $T_C$ with pressure to P=1 kbar in fields $0~<~B~<~11$ T.
A
second-order phase transition, with no volume change, was detected by 
calorimetric
measurements at $T_t$ = 399 K \cite{Meyer,Menyuk}. Between
$T_C$ and $T_t$, the magnetic susceptibility increases with
temperature \cite{Serres,Bates2}, transforming to a
Curie-Weiss law above $T_t$.
X-ray diffraction studies\cite{Wilson} revealed a change in the
crystallographic arrangement at the first-order transition point, from the
NiAs-type ($B8_{1}$) hexagonal structure to the MnP-type ($B31$) orthorhombic one.
The orthorhombic distortion decreases with increasing temperature and
the hexagonal $B31$ structure reappears above $T_t$.
It was soon recognized that a cross-over from high to low spin states
on the manganese
site could occur on going
from the $B8_{1}$ to the $B31$ structure \cite{JBG1} (Fig. \ref{un}). This
conjecture led to a study of
the MnAs$_{1-x}$P$_x$ and MnAs$_{1-x}$Sb$_x$ systems that demonstrated the spin
transition \cite{JBG2}. Studies of the influence of
higher hydrostatic pressure on MnAs led to a modification of the Bean-Rodbell
theory \cite{Menyuk,JBG3}. The spin-state instability appears to be responsible
for a giant magnetoelastic response reported recently by Chernenko {\it et al.}
\cite{Chernenko}.

Our neutron diffraction study of the stability of the
orthorhombic $B31$ phase of MnAs in an applied magnetic field B was 
performed on
the high-resolution powder diffractometer D2B of the Institute Laue-Langevin in
Grenoble, France. Two kinds of polycrystalline samples were measured for this
work, namely a commercial one provided by Western Inorganics and 
another synthetized in
our laboratory by solid-vapor reaction. Mn and As powder (-325 mesh) 
were sealed in
evacuated (P $\simeq$10$^{-5}$ torr.) quartz tubes and heated at 823 K
for 24 hours. This process was repeated several times with intermediate
grindings to obtain the pure compound. In the commercial sample we detected the
presence of MnO impurities, but this has no influence on the structural results
here discussed. For neutron diffraction experiments, two grams of MnAs
were packed in a cylindric vanadium holder and inserted into a 5-tesla
superconducting cryomagnet. The instrument is equipped with 64
collimated $^{3}$He
detectors spaced at 2.5 intervals and covering a 2$\theta$ angular range from
5$^{\circ}$ to 165$^{\circ}$. A complete diffraction pattern was obtained by
moving the detector bank 100 times in angular steps of 
0.025$^{\circ}$. Data have
been collected with incident neutrons of wavelength $\lambda$=1.5943(1) $\AA$
selected by Ge (3 3 5) monochromators.

\begin{figure}
\epsfig{file=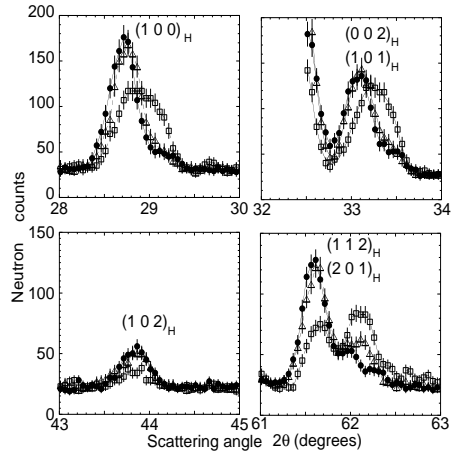,height=6cm,width=6cm}
\caption{Selected regions of the diffraction patterns recorded 
at 27 $^{\circ}$C
with no field (circles), at 49 $^{\circ}$C with no field (squares) and
at 49 $^{\circ}$C with an applied field of 5 tesla (triangles).} 
\label{dous}
\end{figure}

\begin{figure}
\epsfig{file=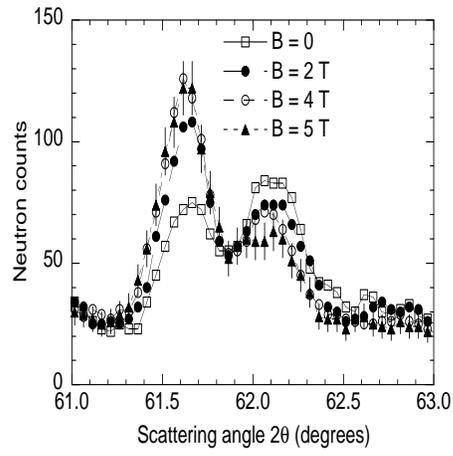,height=6cm,width=6cm}
\caption{Magnetic field dependence of a selected region of the
diffraction pattern
recorded at 322 K, showing the evolution from the orthogonal to the
hexagonal form of MnAs.}
\label{tres}
\end{figure}

At room temperature, below $T_C=$ 313 K, the
diffraction pattern was that of the hexagonal $B8_{1}$ phase with 
lattice parameters
in agreement with those reported by Willis and Rooksby \cite{Willis}. 
Reflections
belonging to the orthorhombic $B31$ structure described by Wilson and Kasper
\cite{Wilson} appeared above $T_C$. As shown in Fig. \ref{dous}, the 
differences
between the diffraction patterns measured in zero field at 300 K and
322 K are readily apparent. However, with the sample held at 322 K,
the application of a magnetic field B modifies substantially the
diffraction profile. For a B= 5 T, the diffraction pattern at 322 K is
essentially the same as that at 300 K in zero applied field. The
hexagonal $B8_{1}$ structure is clearly restored fully by the 
application of a B= 5 T.
As is clear from Fig. \ref{tres}, the extent of the orthorhombic distortion
decreases continuously as B increases, nearly vanishing by B= 4 T. 
These data do
not show a sharp transition, but rather the coexistence of two phases with the
growth of the ferromagnetic phase at the expense of the paramagnetic
$B31$ phase
as B increases. Growth of the ferromagnetic phase is possible because the $B31$
phase is derived from the $B8_{1}$ phase by a co-operative 
displacement of pairs of
[1, -1, 0] rows toward one another to form stronger Mn-Mn bonds, across shared
octahedral-site edges in zig-zag chains within the basal planes. These
displacements also create shorter Mn-As bonds, which raises the
antibonding states that $\sigma$-bond to the As atoms and triggers 
the transition
to a low-spin state \cite{JBG3}.

Extrapolation of the Brillouin temperature dependence of the
magnetization of the $B8_{1}$ phase to above the first-order
transition at $T_{C}$ gives a fictitious Curie temperature
$T_{C}^{*}$ = 388 K.
Since an applied magnetic field B stabilizes the high-spin, ferromagnetic
phase relative to the $B31$ paramagnetic phase, a global
insulator-metal $B31$-$B8_{1}$ transition might be induced by B in the interval
$T_{C}~<~T~<~T_{C}^{*}$. Alternatively, if $B8_{1}$ regions exist 
within a $B31$
matrix in the
interval $T_{C}~<~T~<~T_{t}$, we can expect that below $T_{C}^{*}$ the
ferromagnetic $B8_{1}$
regions will grow in a magnetic field at the expense of the paramagnetic $B31$
matrix to
beyond the percolation threshold. In either case, a CMR would occur 
between $T_{C}$ and $T_{C}^{*}$. If the B field
induces a global $B31$-$B8_{1}$ transition, it would give a CMR 
analogous to that
found, for example, at the metamagnetic transition of the charge and
orbital-ordered CE phase of the perovskite Nd$_{0.5}$Sr$_{0.5}$MnO$_3$
\cite{Shinomura}. If, on the other hand, the magnetic field nucleates and/or
grows the $B8_{1}$ phase, it would give a CMR phenomenon analogous, 
for example,
to that of
the perovskite system (La$_{0.7-x}$Pr$_x$Ca$_{0.3}$)MnO$_3$ \cite{Hwang}.
However, whereas
the CMR occurs in the manganese oxides at too low a temperature to be 
technically
practical, it would occur in MnAs a little above room temperature and
could easily be
adjusted to operate at room temperature by small compositional changes.

\begin{figure}
\epsfig{file=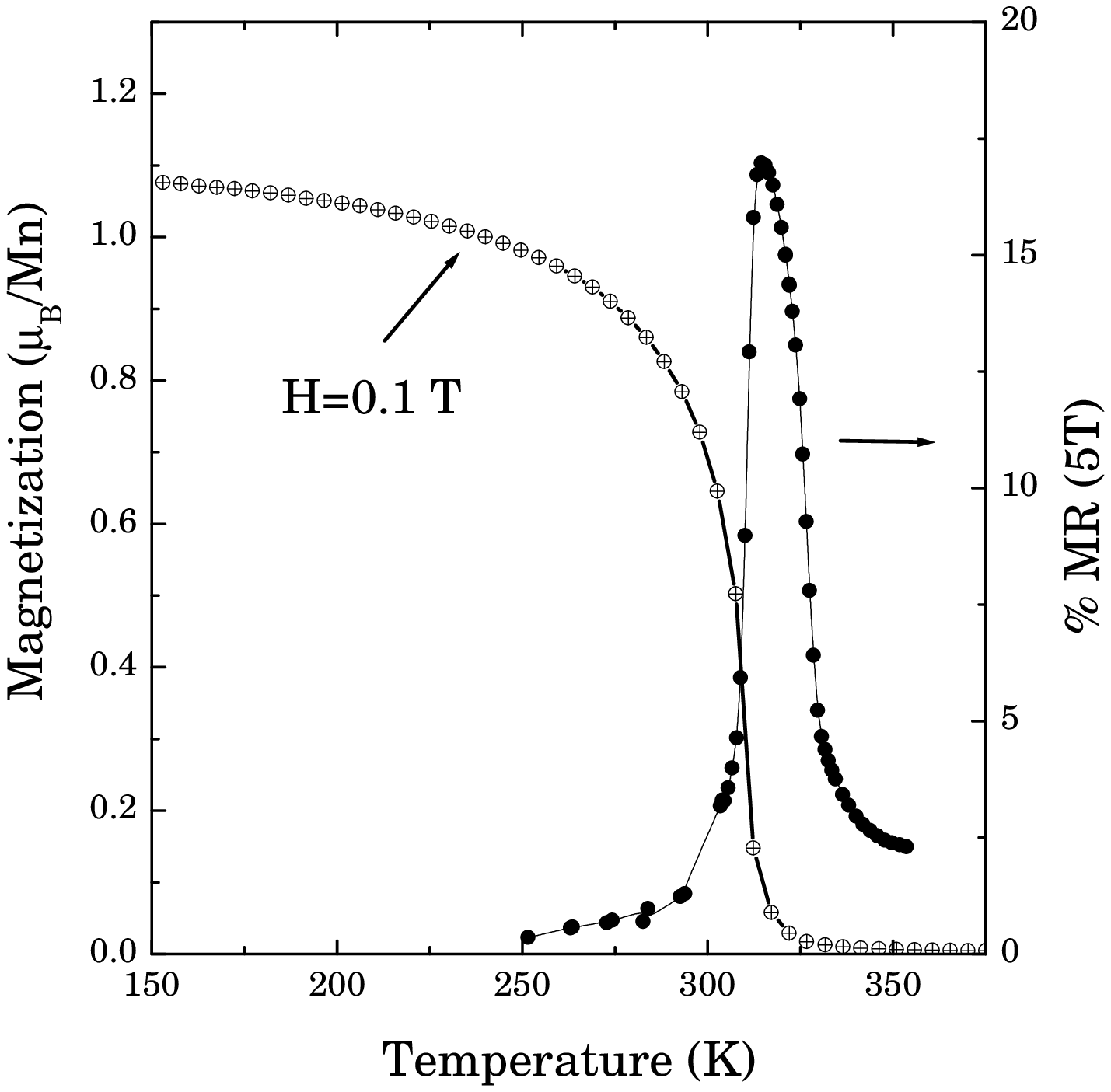,height=6cm,width=6cm}
\caption{Magnetoresistance (MR=[$\rho$(0)-$\rho$(5T)/ $\rho$(0)] x 100) versus
temperature of MnAs and magnetisation measured at 0.1 T in
zero-field-cooling conditions.}
\label{catro}
\end{figure}

Fig. \ref{catro} shows the enhanced magnetoresistance response of MnAs
measured in the predicted temperature
range, $T_{C}~<~T~<~T_{C}^{*}$. Almost the same results were obtained for both
samples
used in this study. Although the absolute value of the magnetoresistance
is not as large as in other compounds like La$_{2/3}$Ca$_{1/3}$MnO$_3$
\cite{Jin}, we call it a CMR to highlight the common origin with that in the
manganites. Analogies between the behavior of MnAs in a magnetic field above
$T_{C}$ and that of the manganese-oxide perovskites are noteworthy. In the
manganese perovskites, the $\sigma$-bonding {\it d} electron of a high-spin
octahedral site Mn(III) ion occupies a twofold-degenerate pair of e-orbitals.
This e-orbital degeneracy is removed by a local distortion of the MnO$_{6/2}$
octahedron; and at lower temperatures and high Mn(III) concentrations, the
local distortions are ordered cooperatively so as to minimize the
associated elastic energy. A recent study of LaMn$_{1-x}$Ga$_x$O$_3$ \cite{Zhou}
has shown that in this single-valent system, dilution of the Mn (III) atoms
suppresses the orbital ordering in local ferromagnetic regions; a vibronic
superexchange is ferromagnetic whereas the orbitally ordered matrix is
antiferromagnetic. In this system, application of B disorders the orbital
order of the matrix to transform a spin glass to a ferromagnet \cite{Zhou}. In
the mixed-valent perovskite system La$_{1-x}$Sr$_x$MnO$_3$, an
orthorhombic-rhombohedral structural phase transition is induced by an external
magnetic field for x=0.17 \cite{Asamitsu}; this transition involves a 
change from
[1,-1,0] to [1,1,1] of the axis of cooperative rotation of the MnO$_{6/2}$
octahedra. A crossover from localized to itinerant behavior of the
$\sigma$-bonding e-electrons occurs in the compositional interval $0.1
\leq x \leq 0.17$ \cite{Liu}. In this interval, hole-rich ferromagnetic regions
are segregated from a hole-poor paramagnetic matrix in the paramagnetic
temperature range $T_C < T \leq 300$ K by cooperative oxygen displacements. The
hole-rich regions are mobile, and they grow in a B to beyond a percolation
threshold by B= 5 T to give the CMR phenomenon. In the interval $0.1 \leq x
\leq 0.15$, the transition at $T_C$ is first-order; and in a narrow temperature
range $T_{OO} < T < T_C$, the orbitals undergo a rearrangement to 
another type of
order below $T_{OO}$. Near x=0.1, the interval $T_{OO} < T < T_C$ narrows, and
the matrix remains orbitally ordered in zero magnetic field to give spin-glass
behavior typical of ferromagnetic clusters having a $T_C$ greater than the
N\'eel temperature of the matrix. However, the orbitals of the matrix become
disordered in a modest B to give ferromagnetic order below a $T_C$. Recently,
Mira {\it et al.} \cite{Mira1} have observed a second-order transition at a
temperature $T^{*}~>~T_C$ in the compositional interval where they had noted a
first-order transition at $T_C$ \cite{Mira2}. In the range $T_C < T < T^{*}$, a
non-Curie-Weiss paramagnetism has been found by de Teresa {\it et al.}
\cite{deTeresa}, similar to the behavior of MnAs in the range $T_C < T < T_t$
\cite{Serres,Guillaud}. Moreo {\it et al.} \cite{Moreo2} have made a
computational study to argue that in the mixed-valent manganese-oxide
perovskites, hole-rich clusters that are metallic and ferromagnetic 
coexist with
a paramagnetic, hole-poor insulating matrix; the ferromagnetic clusters grow in
a magnetic field B. Goodenough \cite{JBG4} has invoked the virial theorem to
make the same deduction for the manganese-oxide perovskites at the crossover
from localized to itinerant behavior of the $\sigma$-bonding electrons.
However, MnAs is single-valent, so the analogy with LaMn$_{1-x}$Ga$_x$O$_3$ is
more appropriate even though this perovskite remains insulating in the
ferromagnetic phase. In MnAs, the basal-plane orbitals on the Mn atoms are
half-filled, and the electrons approach the crossover from localized to
itinerant electronic behavior. In the paramagnetic $B31$ phase, the Mn atoms
become displaced below $T_t$ so as to form stronger Mn-Mn bonds in the zig-zag
chains and weaker Mn-Mn bonds between the chains. These displacements not only
order the Mn-As bonding so as to induce the low-spin state, but they 
also represent
an ordering of the in-plane Mn-Mn bonds that changes the translational symmetry
so as to split the basal-plane band at the Fermi energy. Ferromagnetic
interactions by the other {\it d} electrons are dominant, but these 
interactions
are weakened by the transition to the low-spin state, which reduces the Mn-Mn
separation and strengthens the in-plane Mn-Mn bonding. Since no bonding occurs
between half-filled orbitals with parallel spins, stabilization of 
the high-spin
ferromagnetic $B8_{1}$ phase suppresses in-plane Mn-Mn bonding and 
increases the
in-plane Mn-Mn separation. Stabilization of the ferromagnetic phase in a B with
removal of the in-plane Mn-Mn bonding is to be compared with suppression by a B
of the orbital ordering at the Mn(III) ions in LaMn$_{1-x}$Ga$_x$O$_3$. This is
the first crystallographic transition induced by a magnetic field that involves
breaking of metal-metal bonds rather than orbital disorder.

We are pleased to thank the valuable aid provided by A. Hewat and the
ILL technical staff during the neutron diffraction experiment and the
operation of the cryomagnet in unusual conditions and J. S. Zhou, from the
Texas Materials Institute for his help with the magnetoresistance
measurements. F. R. thanks the Fulbright foundation and MECD from Spain
for a postdoctoral fellowship. We also acknowledge the DGI of the
Ministry of Science and Technology of Spain for financial support
under project MAT2001-3749 JBG thanks the NSF for financial support.

\end{document}